\documentstyle[epsfig]{mn}{}
\begin{document}

\def\mpc{h^{-1} {\rm{Mpc}}}
\def\up{h^{-3} {\rm{Mpc^3}}}
\def\uk{h {\rm{Mpc^{-1}}}}
\def\lsim{\mathrel{\hbox{\rlap{\hbox{\lower4pt\hbox{$\sim$}}}\hbox{$<$}}}}
\def\gsim{\mathrel{\hbox{\rlap{\hbox{\lower4pt\hbox{$\sim$}}}\hbox{$>$}}}}
\def\kms {\rm{km~s^{-1}}}
\def\apj {ApJ}
\def\aj {AJ}
\def\mnras {MNRAS}

\title{Galaxy groups in the 2dF redshift survey: Effects of Environment on
Star Formation}

\author[H.J. Mart\'{\i}nez et al.]
{H.J. Mart\'{\i}nez, A. Zandivarez, M. Dom\'{\i}nguez, M.E. Merch\'an \& D.G. 
Lambas \\
Grupo de Investigaciones en Astronom\'{\i}a Te\'orica y Experimental, 
IATE, Observatorio Astron\'omico, Laprida 854, C\'ordoba, Argentina \\
Consejo de Investigaciones Cient\'{\i}ficas y T\'ecnicas de la Rep\'ublica 
Argentina (CONICET)}
\date{\today}

\maketitle

\begin{abstract}
We estimate the fraction of star forming galaxies in a catalogue of groups, 
constructed from the 2dF galaxy redshift survey by Merch\'an \& Zandivarez. 
We use the $\eta$ spectral type parameter of galaxies and subdivide the sample 
of galaxies in groups into four types depending on the values of the $\eta$ 
parameter following Madgwick et al. 
We obtain a strong correlation between the relative fraction of galaxies 
with high star formation and the parent group virial mass. 
We find that even in the environment of groups with low virial mass 
$M \sim 10^{13} \ M_{\odot}$ the star formation of their member galaxies
is significantly suppressed. The relation between the fraction 
of early-type galaxies and the group virial mass obeys
a simple power law spanning over three orders of magnitude in virial mass. 
Our results show quantitatively the way that the presence of galaxies with 
high star formation rates is inhibited in massive galaxy systems.
\end{abstract}


\section{Introduction} 
Groups of galaxies are one of the most important laboratories in the 
universe to understand how this environment affects the 
galaxy formation. 
One of the main issues on the process of galaxy building is the rate
at which they form stars.
Several works has been devoted to study the star formation rate (hereafter
SFR) in galaxies (Kennicutt 1983, Gallager, Bushouse \& Hunter 1989, 
Kennicutt 1992, Gallagher \& Gibson 1993, Kennicutt, Tamblyn \& Congdom 1994, 
Gallego et al. 1995) and most of them agree that the luminosity of $H_{\alpha}$
provides a direct measure of the global photoionization rate,
which can be used in turn to reliably estimate of SFR in
massive ($M>10 M_{\odot}$) stars. 

Even when most of these studies have been carried out in galaxies,
the extension to systems of galaxies is not very well understood yet 
given the impossibility to account with large samples suitable for 
statistical studies. The effects of environments on the global process of
star formation in galaxy systems is a very important piece on the 
construction of theoretical models of galaxy formation and their 
consequent evolution. 
Some studies have claimed for a strong correlation between group 
early-type galaxy fraction and velocity dispersion. This could result from an 
increase in the early-type fraction and velocity dispersion as a group
evolves, where galaxy morphologies change due to a mechanism such
as mergers or from conditions at the time of galaxy formation 
(Zabludoff 1999). It is possible that mergers cause some evolution
in the early-type fraction of poor groups and cease to be effective
in richer groups and clusters.

Recently, Merch\'an \& Zandivarez (2002) have constructed one 
of the largest sample of 
groups of galaxies until the present using a finding algorithm with the public 
100K data release of the 2dF galaxy redshift survey ($\sim 100000$ galaxies). 
The sample comprise 2209 galaxy groups inside the 2dF angular mask 
with redshifts in the range $0.003 \le z \le 0.25 $. 
The group finding algorithm was designed 
modifying the traditional Huchra \& Geller (1982) finder algorithm in order to 
take into account the 2dF magnitude limit and redshift completeness masks.  

Madgwick et al. (2002) define a new parameter $\eta$ in order to characterise
the galaxy spectra on the 2dF galaxy redshift survey. This parameter is a 
linear combination of the first two projections derived from a Principal 
Component Analysis and their definition is such that its value correlates 
with the strength of absorption-emission features. A negative value of $\eta$ 
is correlated with old stellar populations and strong absorption features 
whereas a positive value is related with young stellar population and strong 
emission lines. Figure 6 of Madgwick et al (2002) shows the strong 
correlation of the $\eta$ parameter  with the equivalent width of $H_{\alpha}$
in emission line galaxies. Consequently, $\eta$ can be interpreted as a 
measure of the current star formation present in each galaxy.

In this work we use the $\eta$ parameter to characterise the star formation
rate of galaxy members of the group catalogue of the 2dF survey. 
We study the correlation of the relative fraction of galaxies with different
values of $\eta$ and the group virial mass.
The outline of this letter is as follows.
In section 2 we describe the group catalogue and the statistical analysis
performed is discussed in section 3. Finally, the main conclusions are given 
in section 4.

\section{The 2dF galaxy group catalogue (2dFGGC)}

Merch\'an \& Zandivarez (2002) identify galaxy groups on the 2dF public 100K 
data release of galaxies with the best redshift estimates within the 
northern (NGP, $-37^{\circ}.5 \leq \delta \leq -22^{\circ}.5$, 
$ 21^h 40^m \leq \alpha \leq 3^h 30^m$) and southern 
($-7^{\circ}.5 \leq \delta \leq 2^{\circ}.5$; $9^h 50^m \leq \alpha 
\leq 14^h 50^m$) strips of the catalogue. 
This sample comprise 84499 galaxies with final $b_j$ magnitudes
corrected for galactic extinction. 
The finder algorithm used on the identification is similar to that developed by 
Huchra \& Geller (1982) but modified in order to take into account the 
sky coverage problems present on the current release of galaxies. 
The redshift completeness, which represent the ratio of the number of galaxies 
for which redshifts have been obtained to the total number of objects 
contained in the parent catalogue, and the magnitude limit mask, which 
correspond with variations of the parent survey magnitude limit with the 
position on the sky, are the two main sky coverage problems on this catalogue
(see Figure 13 and 15 of Colless et al. 2001).

\begin{figure}
\epsfxsize=0.5\textwidth
\hspace*{-0.5cm} \centerline{\epsffile{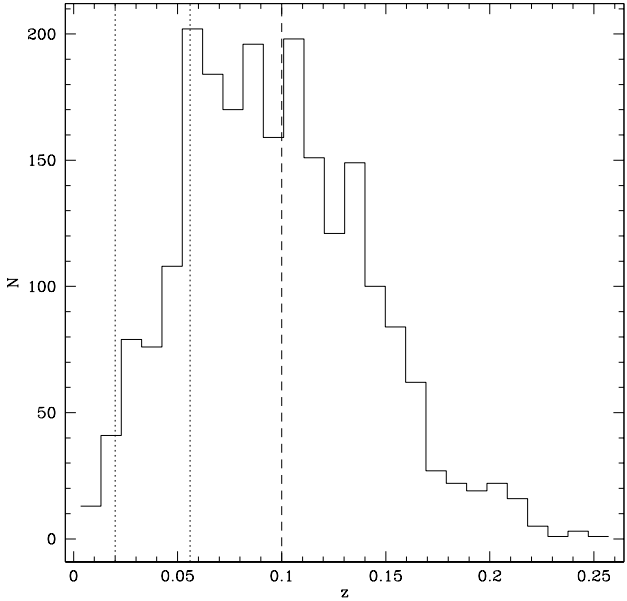}}
\caption{ Redshift distribution of 2dFGGC groups.
Vertical dotted lines are the boundaries of the volume-limited
sample defined in section 3, and dashed line show the upper limit
in redshift of our extended sample.} 
\label{fig1}
\end{figure}

The 2dFGGC was constructed using the values $\delta \rho/\rho=80$ 
and $V_0=200~\kms$ which maximize the group accuracy (see section 4
of Merch\'an \& Zandivarez 2002). 
The resulting groups catalogue comprises a total number
of 2209 galaxy groups with at least 4 members and
mean radial velocities in the range $900 ~\kms\leq V \leq 75000~\kms$.
The limit adopted in the number of 
members in galaxy groups is necessary in order to avoid pseudo-groups.

The virial group masses are estimated using the virial radius and the
velocity dispersion ($M_{vir}=\sigma^2 R_V/G$, Limber \& Mathews 1960) where
the former is computed with the projected virial radius and the later 
with their radial counterpart. 
A robust estimation of this component is obtained applying the biweight 
estimator for groups with richness $N_{tot}\ge 15$ and the gapper estimator 
for poorer groups (Beers, Flynn and Gebhardt 1990, Girardi et al. 1993, 
Girardi and Giuricin 2000).
These methods improve the velocity dispersion estimation in terms of 
efficiency and stability when dealing with small groups. 
Consequently, the catalogue has a mean velocity dispersion of 
$261~ \kms$, a mean virial mass of $8.5\times 10^{13} \ h^{-1} \ 
M_{\odot}$ and a mean virial radius of $1.12 ~ \mpc$.

\section{The $\eta$ spectral type parameter in groups}
In this section the fraction of galaxies of different $\eta$ 
spectral types in 2dFGGC groups is analysed 
as a function of group virial mass.
The $\eta$ parametrization of a galaxy spectral properties
is defined by Madgwick et al (2002) for galaxies in the 100K data release
of the 2dF galaxy redshift survey based upon a Principal Component
Analysis of the galaxy spectra that takes
into account the relative emission/absorption line strength present 
in a galaxy's optical spectrum.
As shown in Madgwick et al (2002), the  equivalent width of $H_{\alpha}$
emission-line, $EW(H_{\alpha})$, is very tightly correlated
to $\eta$ for emission line galaxies.
This classification correlates well with morphology and can be interpreted
as a measure of the relative current star-formation present in each galaxy.
The star formation rate for a galaxy is proportional to $EW(H_{\alpha})$ times
the galaxy luminosity.
Consequently $\eta$ can be used as a measure
of a galaxy star formation rate relative to its luminosity.

We split the galaxies in the 2dFGGC groups 
into the 4 types of Madgwick et al (2002):
\begin{itemize}
\item Type 1: $~~~~~~~~~~\eta < -1.4$,
\item Type 2: $-1.4\leq \eta < ~~1.1$,
\item Type 3: $~~1.1 \leq \eta < ~~3.5$, 
\item Type 4: $~~~~~~~~~~\eta\ge ~~3.5$.
\end{itemize}
This is not a criterion based upon morphology training set but
rather by the shape of the $\eta$-distribution.  

\begin{figure}
\epsfxsize=0.5\textwidth
\hspace*{-0.5cm} \centerline{\epsffile{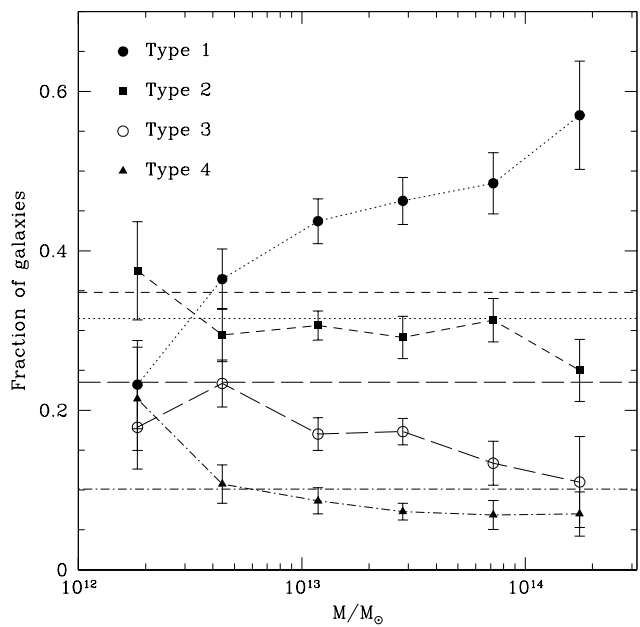}}
\caption{Fraction of the different spectral types as a function
of group virial mass for our volume-limited sample.
Error bars were estimated by the bootstrap resampling technique.
Horizontal lines are the mean fraction of galaxies for
different spectral types within the 2dF Galaxy Redshift Survey
using the same selection criterion than the volume-limited sample.
Dotted line correspond to Type 1 galaxies, short-dashed line to
Type 2, long-dashed to Type 3 and dot-dashed to Type 4.} 
\label{fig1}
\end{figure}

\begin{figure}
\epsfxsize=0.5\textwidth
\hspace*{-0.5cm} \centerline{\epsffile{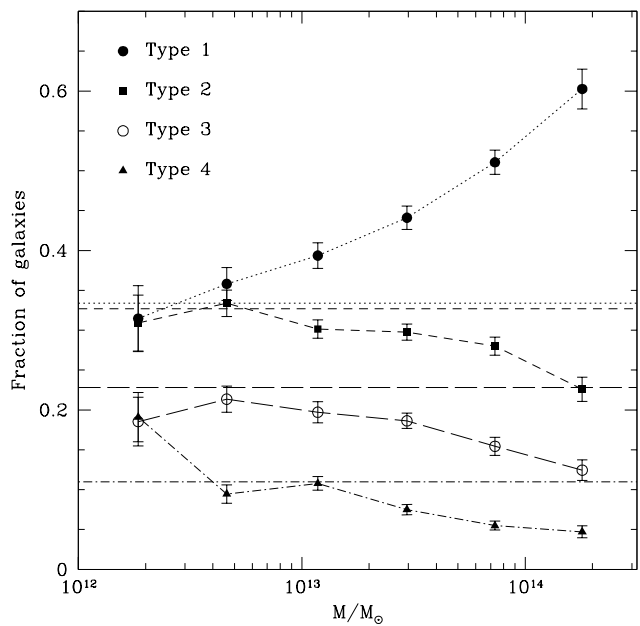}}
\caption{Fraction of the different spectral types as a function
of group virial mass for our extended group sample.
Error bars were estimated by the bootstrap resampling technique.
Horizontal lines are the same as described in Figure 2
but using the current sample restrictions.} 
\label{fig1}
\end{figure}

We have computed the fraction of galaxies of the different types
as a function of group virial mass in two samples; The first one
gives priority to galaxy completeness in order to avoid
significant selection biases, the second sample
is intended to incorporate as many groups as possible trying to avoid
the incompleteness typical of a flux limited sample.

Our first sample of galaxies in groups, which is aimed to
achieve the highest level of completeness, 
was selected taking into account the following prescriptions. Groups 
were limited to the redshift range $0.02\le z \le 0.056$
(see Figure 1).
This range was chosen due to the highly homogeneous distribution
of group virial masses with redshift. Therefore, we prevent 
the possibility of a preferential bias to high mass groups
in the sample. Secondly, we introduce a absolute magnitude cut-off
on the galaxies ($M_{b_J}\lsim -17.2$) attempting a selection which is not 
biased to high luminosity
galaxies, {\em i.e.} no spectral type preference. There are
1522 galaxies in 331 groups in this sample. 

The second group sample considers a wider redshift range, $0.02\le z
\le 0.1$, and the galaxies in these groups have no absolute magnitude
restriction, resulting in 7481 galaxies in 1155 groups.
The large number of groups in this sample allow us to 
increase the reliability of our statistics, but
we have not extended the redshift range beyond $z\sim 0.1$,
since this could introduce strong biases in both, group selection 
and galaxy luminosities.

In Figure 2 we show the fraction of galaxies of  
each spectral type as a function of group virial mass for 
our volume-limited sample.  
We have binned group virial masses into six bins taking 
as abscissa the mean mass of the groups in each interval.
Error bars in Figure 2 were estimated using the bootstrap 
resampling technique. We have also computed the fraction of 
galaxies of each type for all galaxies in the 2dF Galaxy Redshift Survey
with the same selection criterion to determine a comparison level.
It is clear that while the fraction of Type 1 galaxies show
an increasing behavior with group virial mass, the fraction
of galaxies with higher SFR ($\eta \gsim -1.4$) show the opposite trend. 
In particular, for Type 4 galaxies, which have the largest SFR, exists 
a variation of $\sim 75\%$ between the most massive and the less massive groups.
This strong decrease of star forming galaxies on massive groups indicates the 
importance of environmental effects acting on galaxy systems. 

There exists a significant difference between the fraction of galaxies of
Type 1 (galaxies with very low SFR or early type galaxies) in most massive
groups respect to that found when including field galaxies. 
If early-type galaxies are evolved merger remnants as expected in hierarchical
models for galaxy formation, then the galaxy populations of more massive 
groups are more evolved on average. 
Allam et al. (1999) claim that the depressed
star formation in galaxy groups is partly due to a relative over abundance
of early-type galaxies and also to some mechanism that dampens star formation
within late-type spirals. The later suggestion could also explain the
observed decreasing fraction of star forming galaxies ($\eta \gsim -1.4$)
with the virial group mass in our analysis.

The results of the second selected sample are shown in Figure 3.
Even when this sample lacks of the level of confidence of the previous volume 
limited sample, we observe a very good agreement of the trends in
both figures. The increase of the fraction of Type 1 galaxies in the least
massive groups could be due to a possible lack of low luminosity galaxies in the
sample. This effects could also explain the variations in the mean global
fractions between Figure 2 and 3.

Regarding the contribution of each galaxy spectral type to the current SFR
in groups, the influence of Type 1 galaxies is negligible
since they are dominated by old stellar populations.
Type 2 galaxies dominate in number over 
the remaining star forming types independently
of group virial mass. However, since Type 3 galaxies
have larger values of $EW(H_{\alpha})$ and similar luminosities, 
their contribution to the SFR could be even higher than 
that of Type 2 galaxies.
On the other hand, Type 4 corresponds to the tail of 
the $\eta$ distribution (see 
figure 4 of Madwick et. al 2002) and is dominated by particularly
active galaxies such as starbursts and AGNs. 
Consequently special care should be taken when computing the contribution
of these galaxies to the SFR, specially in low mass groups where
the fraction of Type 4 galaxies is more important.

\section{CONCLUSIONS}
We have analysed the fraction of star forming galaxies in 
environments corresponding to groups and poor clusters of galaxies. 
The 2dFGGC is the largest sample available at the present
and can be easily divided several times whilst still maintaining
very reliable statistics. 
Our subsamples are the largest data sets used in the calculation
of galaxy populations in groups so far.

The results of the analysis reported in this work clearly show a continuous
trend of decreasing fraction of spectral types associated to star forming
galaxies with group virial mass.
Groups with virial mass $M\sim 2-4 \times 10^{12} \ M_{\odot}$ have a
relative fraction of galaxy spectral types similar to the global values.
These results imply that 
even the environment of low mass systems 
$M \sim 10^{13} \ M_{\odot}$ is effective in diminishing the process
of star formation in their member galaxies.

It is interest to note the way that the fraction of early-type galaxies 
(Type 1) increases monotonically 
with group virial mass which can be well fitted by a power-law of the form 
$\log(F_{Type 1})=0.14\log(M/M_{\odot})-2.2$. This simple law  provides a 
suitable fit to the fraction of Type 1 galaxies spanning over 
three orders of magnitude in group virial masses. 

Our tests with two redshift restricted 
samples provide firm evidence of the lack of 
biases in our results that could arise due to the different galaxy 
luminosity functions of each spectral type as well as possible biased 
selection of high mass groups.

Recently Lewis et al. (2002) have studied a sample of seventeen known 
galaxy clusters using 2dFGRS spectra to compute SFR as function of
local galaxy density and cluster-centric radius. They have found that
the dependence of SFR on local density is independent of cluster velocity 
dispersion and presumably mass. 
It should be taken into account the different characteristics of the sample 
analysed in that paper and those analysed in this work.
The sample study by Lewis et al. (2002) comprises more massive 
systems than the groups in our samples. 
A detailed analysis of the dependence of spectral types fractions 
on local galaxy density and group-centric distance for our samples,
is given in Dom\'\i nguez et al. (2002). They find that 
the more massive groups show a significant dependence of the
fraction of low star forming galaxy on local galaxy density and
group-centric radius whereas groups with lower masses show no
significant trends. 
As we have discussed above, it is important to remark that 
our results and those obtained by Dom\'\i nguez
et al. (2002) refer to relative fractions of galaxy spectral types and not
to current SFR in groups.

The present results could be very useful to  provide constraints to
the theoretical models of galaxy formation and evolution. 
In particular, the statistical behaviors reported
in this work should be taken into account when considering the influence 
of environment on galaxy formation and evolution in semi-analytical models.

\section*{Acknowledgments}
We thank the referee Jaime Zamorano for helpful comments.
We thank to Peder Norberg and Shaun Cole for kindly providing the 
software describing the mask of the 2dFGRS and to the 2dFGRS Team
for having made available the actual data sets of the sample.
We also thank Hernan Muriel for reading the manuscript and useful
comments.
This work has been partially supported by Consejo de Investigaciones 
Cient\'{\i}ficas y T\'ecnicas de la Rep\'ublica Argentina (CONICET), the
Secretar\'{\i}a de Ciencia y T\'ecnica de la Universidad Nacional de C\'ordoba
(SeCyT) and Fundaci\'on Antorchas, Argentina.


\begin{thebibliography}{}

\bibitem[Allam et al. (1999)]{allam1999}
Allam S., Tucker D., Lin H., Hashimoto Y., 1999, \apj, 522L, 89. 

\bibitem[Beers et al. (1990)]{beers1990}
Beers T. C., Flynn K., Gebhardt K., 1990, \aj, 100, 32. 

\bibitem[Colless et al. (2001)]{colles01}
Colless M., et al. (2dFGRS Team), 2001, \mnras, 328, 1039.

\bibitem[De Propris et al. (2002)]{deproris02}
De Propris R., et al.(2dFGRS Team), 2002, \mnras, 329, 87.

\bibitem[Dom\'\i nguez et al. (2002)]{mardom}
Dom\'\i nguez M., Zandivarez A., Mart\'\i nez H., Merch\'an M., Muriel H. \&
Lambas D. G., 2002, MNRAS, submitted.

\bibitem[Girardi et al.(1993)]{girardi93}
Girardi M., Biviano A., Giuricin G., Mardirossian F. \& Mezzetti M., 1993, 
\apj, 404, 38.

\bibitem[Gallagher et al.(1989)]{gallagher89}
Gallagher J.S., Bushouse H., Hunter D.A., 1989, \aj, 97, 700.

\bibitem[Gallagher et al.(1993)]{gallagher93}
Gallagher J.S., Gibson S.J., 1993, in Panchromatic View of Galaxies, ed. G.
Hensler, C. Theis \& J.S. Gallagher (Gif-sur-Yvette: Editions Fronti\`eres),
207.

\bibitem[Gallego et al.(1995)]{gallego95}
Gallego J., Zamorano J., Arag\'on-Salamanca A., Rego M., 1995, \apj, 455, L1.

\bibitem[Girardi et al.(2000)]{girardi00}
Girardi M., Giuricin G., 2000, \apj, 540, 45.

\bibitem[Huchra \& Geller (1982)]{huchra82}
Huchra J.P., Geller M.J., 1982, \apj, 257, 423.

\bibitem[Kenicutt (1983)]{kenicutt83}
Kenicutt R.C., 1983, \apj, 272, 54.

\bibitem[Kenicutt (1992)]{kenicutt92}
Kenicutt R.C., 1992, \apj, 388, 310.

\bibitem[Kenicutt et al. (1994)]{kenicutt94}
Kenicutt R.C., Tamblyn P., Congdon C.W., 1994, \apj, 435, 22.

\bibitem[Lewis et al. (2002)]{lewis}
Lewis I., et al. (2dFGRS Team), 2002, \mnras, submitted (astro-ph/0203336). 

\bibitem[Limber \& Mathews (1960)]{limber60}
Limber D.N., Mathews W.G.,1960, \apj, 132, 286. 

\bibitem[Madgwick et al. (2002)]{madgwick01}
Madgwick D., et al. (2dFGRS Team), 2002, \mnras, submitted (astro-ph/0107197).

\bibitem[Merchan \& Zandivarez (2002)]{merchan00}
Merch\'an M.E., Zandivarez A., 2002, \mnras, submitted.

\bibitem[Norberg et al. (2001)]{norberg01}
Norberg P., et al. (2dFGRS Team), 2001, \mnras, 328, 64.

\bibitem[Percival et al. (2001)]{percival01}
Percival W.J., et al. (2dFGRS Team), 2001, \mnras, 327, 1297.

\bibitem[Zabludoff (1999)]{zablu}
Zabludoff A., 1999, IAUS, 192, 433.

\end{thebibliography}
\end{document}